\documentclass[10pt,letterpaper]{article}
\usepackage{opex3}
\usepackage{color}
\usepackage{xcolor}
\usepackage{cite}
\usepackage{amsmath,amssymb,amsthm}
\usepackage[colorlinks=true,urlcolor=blue,linkcolor=black,citecolor=black,breaklinks=true,bookmarks=false]{hyperref}

\newcommand{\mat}[4]{\left[\begin{array}{cc} #1 & #2 \\ #3 &  #4  \end{array}\right]}

\begin{document}

\title{Quantum optics of lossy asymmetric beam splitters}

\author{Ravitej Uppu,$^{1,*}$ Tom A. W. Wolterink,$^{1,2}$ Tristan B. H. Tentrup,$^1$ and Pepijn W. H. Pinkse$^{1,\dagger}$}
\address{$^1$Complex Photonic Systems (COPS), MESA+ Institute for Nanotechnology, University of Twente, P.O. Box 217, 7500 AE Enschede, The Netherlands\\ $^2$ Laser Physics and Nonlinear Optics (LPNO), MESA+ Institute for Nanotechnology, University of Twente, P.O. Box 217, 7500 AE Enschede, The Netherlands\\$^*$r.uppu@utwente.nl\\$^\dagger$p.w.h.pinkse@utwente.nl}


\begin{abstract}
We theoretically investigate quantum interference of two single photons at a lossy asymmetric beam splitter, the most general passive 2$\times$2 optical circuit. The losses in the circuit result in a non-unitary scattering matrix with a non-trivial set of constraints on the elements of the scattering matrix. Our analysis using the noise operator formalism shows that the loss allows tunability of quantum interference to an extent not possible with a lossless beam splitter. Our theoretical studies support the experimental demonstrations of programmable quantum interference in highly multimodal systems such as opaque scattering media and multimode fibers.
\end{abstract}


\section{Introduction}
Multiphoton quantum correlations are crucial for quantum information processing and quantum communication protocols in linear optical networks \cite{Knill2001,Kok2007}. Beam splitters form a fundamental component in the implementation of these linear optical networks \cite{Reck1994}. They have been realized in a variety of systems including integrated optics, atomic systems, scattering media, multimode fibers, superconducting circuits and plasmonic metamaterials \cite{Politi2008,Lopes2015,Wolterink2016,Defienne2016,Petta2010,Lang2013,Heeres2013,Fakonas2014}. In plasmonic systems, beam splitters have been used to generate coherent perfect absorption in the single-photon regime \cite{Roger2015,Baldacci2015} and on-chip two-plasmon interference \cite{Heeres2013, Fakonas2014}. Inherent losses in optical systems are unavoidable and can arise from dispersive ohmic losses or from imperfect control and collection of light in dielectric scattering media. The effect of losses in beam splitters has attracted a lot of theoretical attention due to the fundamental implications of unavoidable dispersion in dielectric media \cite{Barnett1998, Gruner1996, Jeffers2000, Scheel2000}. However, all these studies have dealt with either symmetric (equal reflection-transmission amplitudes for both input arms) or balanced (equal reflection and transmission amplitudes in each arm) beam splitters. In this article, we analyze the most general two-port beam splitter which can be lossy, asymmetric and unbalanced, and find the non-trivial constraints on the matrix elements. We derive general expressions for the probabilities to measure zero, one or two photons in the two outputs when a single photon is injected in each of the two inputs. Further, we comment on the possible measurements of quantum interference through coincidence detection in a Hong-Ou-Mandel-like setup \cite{Hong1987}.

\begin{figure}[ht]
\centering
\includegraphics[width=13cm]{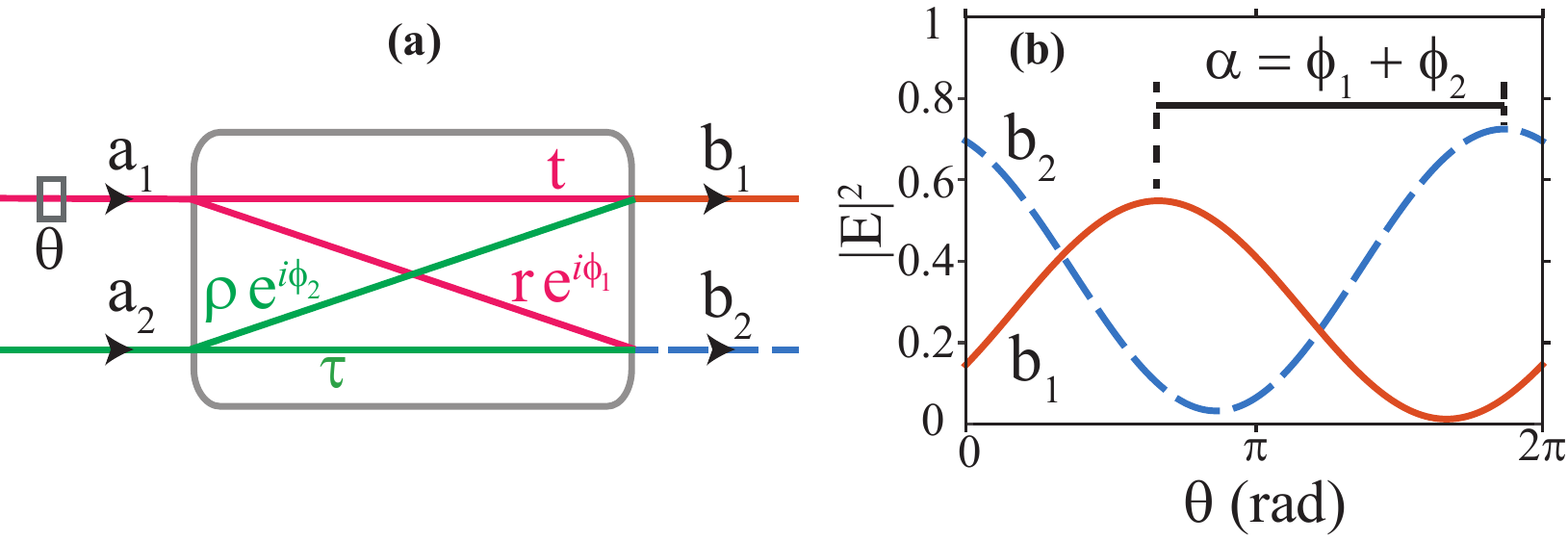}
\caption{(a) depicts the schematic of a general 2$\times$2 beam splitter with input ports $a_1$ and $a_2$ and output ports $b_1$ and $b_2$. The transmission-reflection amplitudes for light in input ports $a_1$ and $a_2$ are $t$-$r$ and $\tau$-$\rho$ respectively. (b) illustrates the output power at ports $b_1$ (orange curve) and $b_2$ (blue curve) as phase $\theta$ is varied between 0 and 2$\pi$ at input port $a_1$. The phase difference between the peak amplitudes is $\alpha = \phi_1 + \phi_2$.}
\label{fig:intro}
\end{figure}

A general two-port beam splitter or a linear optical network consists of two input ports $a_1, a_2$ and two output ports $b_1, b_2$ as schematized in Fig. \ref{fig:intro}(a). The linearity of the beam splitter gives rise to a linear relation between the electric fields, $E(b_i) = \sum_{i,j} s_{ij} E(a_j)$. The complex numbers $s_{ij}$ are the elements of a scattering matrix $S$ and correspond to the transmission and reflection coefficients with $s_{11} = t \exp i\phi_{11}$, $s_{22} = \tau \exp i\phi_{22}$, $s_{12} = \rho \exp i\phi_{12}$, and $s_{21} = r \exp i\phi_{21}$, where $t,\tau,r,\rho$ are positive real numbers. The phases $\phi_{ij}$ are not all independent and can be reduced to $\phi_1$ and $\phi_2$ which correspond to the phase differences between transmission and reflection at a given input port. This gives the scattering matrix
\begin{equation}
S = \mat{t}{\rho e^{i \phi_2}}{r e^{i \phi_1}}{\tau}.
\label{eq:smat}
\end{equation}
Without further constraints on the matrix elements, the scattering matrix $S$ need not be unitary. Special cases include the balanced beam splitter where $\tau = \rho; t = r$ and the symmetric beam splitter where $\tau = t; \rho \exp(i\phi_2)= r \exp(i\phi_1)$. 

The six parameters in the scattering matrix are required to describe the behavior of the output intensities. Fig. \ref{fig:intro}(b) illustrates the intensities $|E|^2$ at $b_1$ and $b_2$ as the phase of the input coherent field at $a_1$ is varied (with phase at $a_2$ fixed). For a general beam splitter, the amplitudes, intensity offsets and phase offsets at the two output ports can be completely free. Of particular interest is the value of the phase difference $\alpha$, which, as we discuss in the subsequent sections, determines the visibility of quantum interference between two single photons.

\section{Energy constraints}
The beam-splitter scattering matrix in Eq. (\ref{eq:smat}) is defined without any constraints on the parameters. However, the physical constraint that the output energy must be less than or equal to the input energy imposes restrictions on the parameters as derived below. Let us consider the scenario where coherent states of light with fields $E_1$ and $E_2$ are incident at input ports $a_1$ and $a_2$ respectively. Energy conservation at a lossy beam splitter imposes the restriction that the total output powers in the arms should be less than or equal to the input,
\begin{equation}
|t E_1 + \rho e^{i \phi_2} E_2 |^2  + |r E_1 e^{i \phi_1} + \tau E_2 |^2 \leq |E_1|^2 + |E_2|^2 .
\end{equation}
The two input coherent state fields can be related through a complex number $c = |c| e^{-i \delta}$ as $E_2 = c E_1$, which gives
\begin{equation}
t \rho  \cos (\phi_2 - \delta) + \tau r \cos (\phi_1 + \delta) \leq \frac{(1 - t^2 - r^2) + |c|^2 (1 - \tau^2 - \rho^2)}{2 |c|}.
\label{eq:3eq}
\end{equation}

As the inequality holds for all values of $|c|$, it should also hold in the limiting case where the right hand side of Eq. (\ref{eq:3eq}) is minimized. This occurs for $|c|^2 = (1-t^2-r^2)/(1-\tau^2-\rho^2)$. Upon substitution, the inequality becomes
\begin{equation}
t \rho \cos(\phi_2 - \delta) + \tau r \cos (\phi_1 + \delta) \leq \sqrt{(1-t^2-r^2)(1-\tau^2-\rho^2)}.
\end{equation}
The above inequality can be algebraically manipulated using trigonometric identities into the following form
\begin{equation}
\sqrt{t^2 \rho^2 + \tau^2 r^2 + 2 \tau \rho r t \cos(\phi_1 + \phi_2)}\sin (\delta + \theta_\textrm{off})  \leq \sqrt{(1-t^2-r^2)(1-\tau^2-\rho^2)},
\end{equation}
where $\theta_{\textrm{off}} = \arctan [(t \rho \cos \phi_2  + \tau r \cos \phi_1)/(t \rho \sin \phi_2 - \tau r \sin \phi_1)]$. As the inequality holds for all values of $\delta$, it should hold in the limiting case of the maximum value of the left hand side which occurs when $\delta + \theta_{\textrm{off}} = \pi/2$. Substituting $\alpha = \phi_1 + \phi_2$ results in the following inequality in terms of the reflection and transmission amplitudes
\begin{equation}
\sqrt{t^2 \rho^2 + \tau^2 r^2 + 2 \tau \rho r t \cos \alpha} \leq \sqrt{(1-t^2-r^2)(1-\tau^2-\rho^2)}.
\label{eq:6param}
\end{equation}
For the lossless beam splitter, the equality results in $\alpha = \pi$. For a symmetric balanced beam splitter, i.e. $t=r=\tau=\rho$ and $\phi_1 = \phi_2$, Eq. (\ref{eq:smat}) reduces to the well-known beam splitter matrix \cite{Gerry2005}
\begin{equation}
S_\textrm{sym-bal} = t \mat{1}{i}{i}{1}
\end{equation}

\begin{figure}[ht]
\centering
\includegraphics[width=8cm]{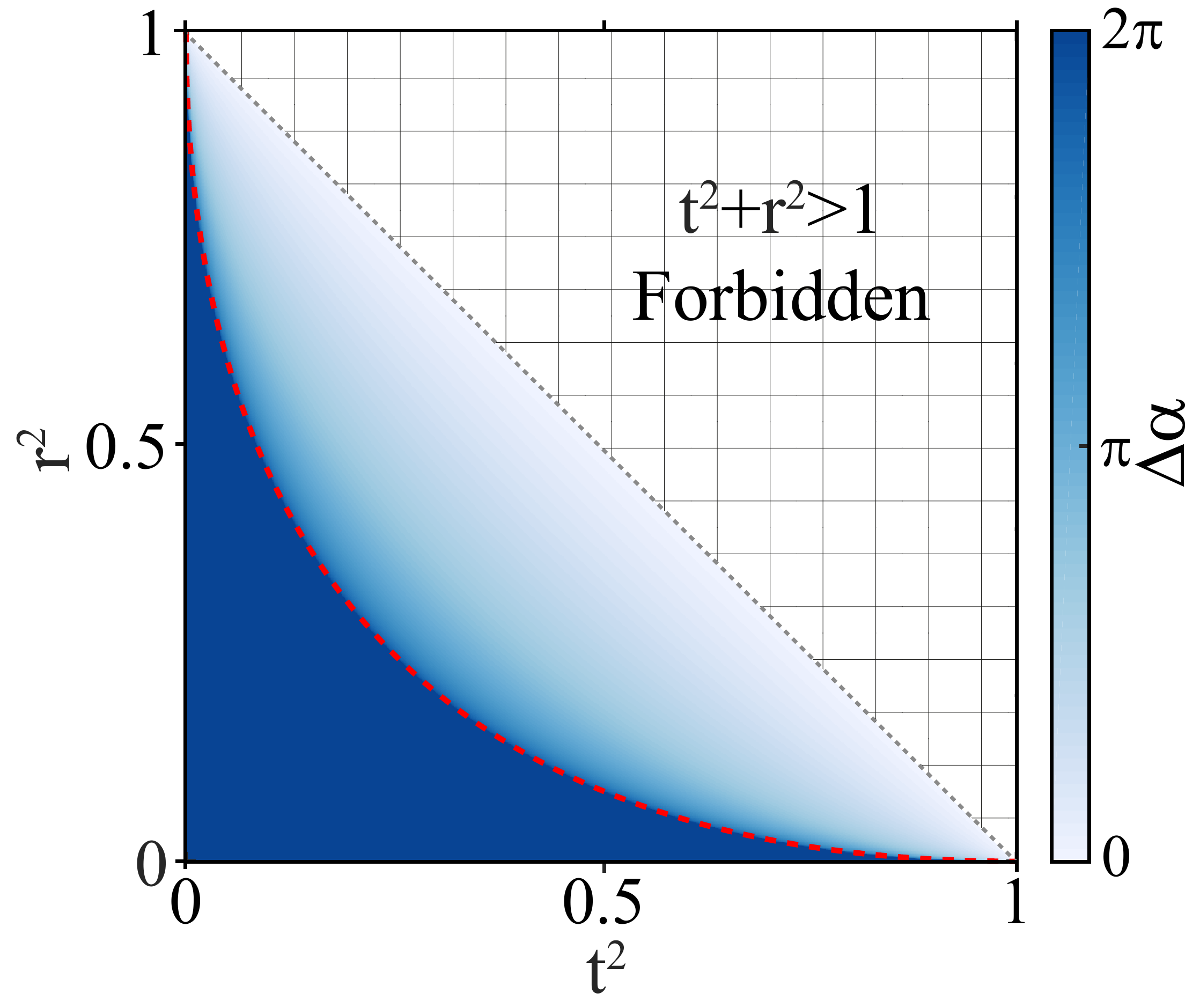}
\caption{The figure depicts the allowed tunable width $\Delta \alpha$ around $\pi$. The anti-diagonal line ($r^2+t^2=1$) separating the allowed from the forbidden region corresponds to lossless beam splitter. The red dashed line is the curve $t + r = 1$. Any lossy circuit that satisfies $t + r \leq 1$ allows complete tunability of $\alpha \in [0, 2 \pi]$.}
\label{fig:all-alp}
\end{figure}

The inequality in equation (\ref{eq:6param}) corresponds to the most general constraint on the parameters of a passive lossy asymmetric beam splitter. For the sake of clarity, we will discuss the case of a lossy symmetric beam splitter with $\tau = t$ and $\rho = r$. In this scenario, the inequality has three parameters
\begin{equation}
\left\lvert \cos \frac{\alpha}{2} \right\rvert \leq  \frac{1-t^2-r^2}{2tr}.
\end{equation}
This inequality results in an allowed range of $\alpha$ between $[\pi - \frac{\Delta \alpha}{2}, \pi + \frac{\Delta \alpha}{2}]$. Fig. \ref{fig:all-alp} depicts the tuning width $\Delta\alpha$ as a function of reflectance $r^2$ and transmittance $t^2$. The lossless beam splitters lie on the diagonal line that separates the forbidden and allowed regions. Evidently, lossless beam splitters have $\Delta \alpha$ = 0, i.e. the phase difference $\alpha$ between the output arms is fixed and equals $\pi$. With increasing losses in the beam splitter, $\Delta \alpha$ increases and achieves a maximum value of 2$\pi$, i.e. complete tunability of $\alpha$. The beam splitters that exactly satisfy $t+r = 1$ (red dotted line) correspond to those lossy beam splitters that allow completely programmable operation with maximum transmission or reflection. In the following section, we discuss the effect of this tunability on the quantum interference between two single photons incident at the input ports of the general beam splitter.

\section{Quantum interference of two single photons}
The quantum-mechanical input-output relation of the lossy asymmetric beam splitter can be written using the scattering matrix in Eq. (\ref{eq:smat}). From this point, we explicitly take into account the frequency dependence that is required to calculate the Hong-Ou-Mandel interference between single photons incident at the input ports. 
\begin{equation}\label{eq:matrix}
\left[\begin{array}{c} 
\hat{b}_{1}(\omega) \\
\hat{b}_{2}(\omega) \\
\end{array}\right]
=\mat{t(\omega)}{\rho(\omega) e^{i \phi_2}}{r(\omega) e^{i \phi_1}}{\tau(\omega)} 
\left[\begin{array}{c} 
\hat{a}_{1}(\omega) \\
\hat{a}_{2}(\omega) \\
\end{array}\right] + \ \left[\begin{array}{c} 
\hat{F}_{1}(\omega) \\
\hat{F}_{2}(\omega) \\
\end{array}\right].
\end{equation}
The operators $\hat{a}_i (\omega)$ and $\hat{b}_i (\omega)$ are creation-annihilation operators of photons at the input and output ports, respectively. The canonical commutation relations of these operators are satisfied even in the presence of loss.
\begin{align}
[ \hat{a}_i(\omega), \hat{a}_j(\omega') ] &= 0; &\forall i,j \in \{1,2\}, \\
[ \hat{a}_i(\omega), \hat{a}_j ^\dagger (\omega')] &= \delta_{ij} \delta(\omega-\omega'); &\forall i,j \in \{1,2\}, \\ 
[ \hat{b}_i(\omega), \hat{b}_j(\omega') ] &= 0; &\forall i,j \in \{1,2\}, \\
[ \hat{b}_i(\omega), \hat{b}_j ^\dagger (\omega')] &= \delta_{ij} \delta(\omega-\omega'); &\forall i,j \in \{1,2\}.
\end{align}
The introduction of noise operators $\hat{F}_i (\omega)$ in Eq. (\ref{eq:matrix}), which represent quantum fluctuations, are necessary in the presence of loss as reported earlier \cite{Barnett1998,Huttner1992,Zmuidzinas2003}. We assume that the underlying noise process is Gaussian and uncorrelated across frequencies. 

The commutation relations of the noise operators can be calculated as the noise sources are independent of the input light, i.e. 
\begin{align}
[ \hat{a}_i (\omega), \hat{F}_j (\omega')] = [ \hat{a}_i (\omega), \hat{F}_j ^\dagger (\omega')]  = 0; \hspace{0.5 cm} \forall i,j \in \{1,2\}, 
\end{align}
which results in
\begin{align}
[ \hat{F}_i (\omega), \hat{F}_j (\omega') ] &= [ \hat{F}_i ^\dagger (\omega), \hat{F}_j ^\dagger (\omega')]  = 0; \hspace{0.5 cm} \forall i,j \in \{1,2\},\\
[ \hat{F}_1 (\omega), \hat{F}_1 ^\dagger (\omega')] &= \delta(\omega-\omega')[1 - t^2(\omega) - \rho^2(\omega)], \\
[ \hat{F}_2 (\omega), \hat{F}_2 ^\dagger (\omega')] &= \delta(\omega-\omega')[1 - \tau^2(\omega) - r^2(\omega)], \\
[ \hat{F}_1 (\omega), \hat{F}_2 ^\dagger (\omega')] &= -\delta(\omega-\omega')[t(\omega) r(\omega) e^{-i \phi_1} + \rho (\omega) \tau (\omega) e^{i \phi_2}],\\
[ \hat{F}_2 (\omega), \hat{F}_1 ^\dagger (\omega')] &= -\delta(\omega-\omega')[t(\omega) r(\omega) e^{i \phi_1} + \rho (\omega) \tau (\omega) e^{-i \phi_2}].
\end{align}

To calculate the effect of the quantum interference, let us suppose that a single photon with frequency $\omega_1$ is incident at input $a_1$ and another single photon with frequency $\omega_2$ is incident at input $a_2$. The two photons together have a bi-photon amplitude $\psi(\omega_1,\omega_2)$ which results in the following input state,
\begin{equation}
\lvert \Psi \rangle = \lvert 1_1, 1_2 \rangle = \int _0 ^\infty  d\omega_1 \int _0 ^\infty d \omega_2 \psi(\omega_1,\omega_2) \hat{a}_1 ^\dagger (\omega_1) \hat{a}_2 ^\dagger (\omega_2) \lvert 0 \rangle.
\end{equation}
The bi-photon amplitude $\psi(\omega_1,\omega_2)$ is normalized as $\int_0 ^\infty d \omega_1 \int _0 ^ \infty d \omega_2 \lvert \psi(\omega_1,\omega_2) \rvert^2 = 1$, ensuring that the state vector $\lvert \Psi \rangle$ is normalized.

In a lossy beam splitter, there are in total six possible outcomes with either two, one or zero photons at each output port. The probabilities of these outcomes can be represented as expectation values of the number operators for the output ports, defined as 
\begin{equation}
\hat{N}_i (\omega) =\int _0 ^\infty d \omega \hat{b}_i ^\dagger (\omega) \hat{b}_i (\omega) \hspace{0.5cm}  i \in\{1,2\}.
\end{equation} 
Assuming that detectors have perfect efficiency, the probabilities can be calculated using the Kelley-Kleiner counting formulae \cite{Kelley1964} and can be grouped into 3 sets:
\begin{itemize}
	\item No photon lost
	\begin{align}
	P(2_1,0_2) &= \frac{1}{2} \langle \hat{N}_1   (\hat{N}_1   - 1) \rangle , \\
	P(0_1,2_2) &= \frac{1}{2} \langle \hat{N}_2 (\hat{N}_2   - 1) \rangle , \\
	P(1_1,1_2) &= \langle \hat{N}_1   \hat{N}_2   \rangle
	\end{align}
	\item One photon lost
	\begin{align}
	P(1_1,0_2) &= \langle \hat{N}_1   \rangle - \langle \hat{N}_1   (\hat{N}_1   - 1) \rangle - \langle \hat{N}_1   \hat{N}_2   \rangle , \\
	P(0_1,1_2) &= \langle \hat{N}_2   \rangle - \langle \hat{N}_2   (\hat{N}_2   - 1) \rangle - \langle \hat{N}_1   \hat{N}_2   \rangle
	\end{align}
	\item Both photons lost
	\begin{equation}
	P(0_1,0_2) = 1 - \langle \hat{N}_1   \rangle - \langle \hat{N}_2   \rangle + \langle \hat{N}_1   \hat{N}_2   \rangle + \frac{1}{2} \langle \hat{N}_1   (\hat{N}_1  - 1) \rangle + \frac{1}{2} \langle \hat{N}_2   (\hat{N}_2  - 1) \rangle
	\end{equation}	
\end{itemize}
Of particular interest is the coincidence probability $P(1_1,1_2)$ which decreases to zero at a lossless, symmetric balanced beam splitter, which is known as the Hong-Ou-Mandel effect \cite{Hong1987}.

Under the assumption that coefficients $t, r, \tau, \rho$ are frequency independent, the expectation values of the number operators are 
\begin{align}
\langle \hat{N}_1 \rangle &= t^2 + \rho^2,\\
\langle \hat{N}_2 \rangle &= \tau^2 + r^2 ,\\
\langle \hat{N}_1 (\hat{N}_1 -1) \rangle &= 2 t^2 \rho^2 [1 + I_{\textrm{overlap}}(\delta t)], \\
\langle \hat{N}_1 (\hat{N}_1 -1) \rangle &= 2 \tau^2 r^2 [1 + I_{\textrm{overlap}}(\delta t)] ,\\
\langle \hat{N}_1 \hat{N}_2 \rangle &= t^2 \tau^2 + r^2 \rho^2 + 2 \tau \rho t r I_{\textrm{overlap}}(\delta t) \cos \alpha ,
\end{align}
where $I_{\textrm{overlap}}(\delta t)$ is the spectral overlap integral of the two single photons at the input ports of the beam splitter, given as
\begin{equation}
I_{\textrm{overlap}}(\delta t) = \int _0 ^\infty d\omega_1 \int _0 ^\infty d\omega_2 \psi(\omega_1,\omega_2) \psi^* (\omega_2,\omega_1) \exp[-i(\omega_1 - \omega_2)\delta t].
\end{equation}
Usually, in experimental measurements of quantum interference, the time delay is varied to retrieve the Hong-Ou-Mandel dip in the coincidence rates.

\begin{figure}[ht]
\centering
\includegraphics[width=13cm]{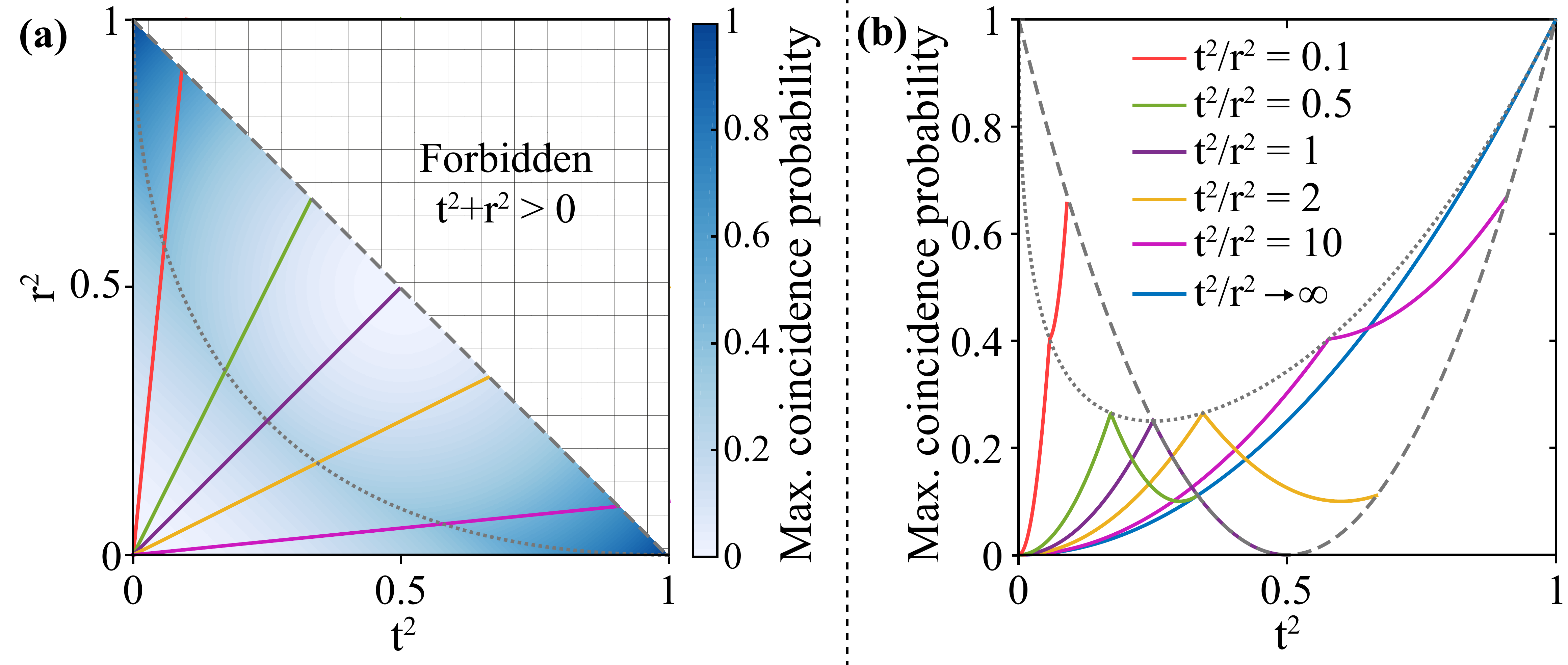}
\caption{(a) The variation of the maximal coincidence rate $\textrm{max}_\alpha P(1_1,1_2)$ in a general beam splitter is shown as a function of reflectance and transmittance. The solid curves in (a) and (b) correspond to cross-sections along different imbalance values $t^2/r^2$. The dashed curve in (a) and (b) is the coincidence probability in a lossless beam splitter. The dotted curve in (a) and (b) depicts the coincidence probability of beam splitters with $t + r$ = 1.}
\label{fig:maxp11}
\end{figure}

For the case of a symmetric beam splitter as discussed in Fig. \ref{fig:all-alp}, the probabilities of different outcomes are
\begin{align}
P(1_1,1_2) &= t^4 + r^4 + 2 t^2 r^2 I_\textrm{overlap}(\delta t) \cos \alpha ,\\
P(2_1,0_2) &= P(0_1,2_2) = t^2 r^2 [1 + I_\textrm{overlap}(\delta t)] ,\\
P(1_1,0_2) &= P(0_1,1_2) = t^2 + r^2 - t^4 - r^4 -2t^2 r^2 \{1 + I_\textrm{overlap}(\delta t)[1+\cos \alpha] \},\\
P(0_1,0_2) &= 1 - 2(t^2+r^2) + t^4 + r^4 + 2 t^2 r^2 \{1+ I_\textrm{overlap} (\delta t)[1 + \cos \alpha]\}.
\end{align}
The coincidence probability $P(1_1,1_2)$ varies sinusoidally with $\alpha$. For a lossless and balanced beamsplitter, $\alpha$ = $\pi$ and the coincidence probability is zero, corresponding to the well-known Hong-Ou-Mandel bunching of photons. However in a lossy beam splitter, the coincidence probability between perfectly indistinguishable photons varies with $\alpha$ from $(t^2 - r^2)^2$ to $(t^2 + r^2)^2$, assuming $\Delta \alpha$ = 2$\pi$. Further, it is interesting to note that the probability of photon bunching at the first output port, $P(2_1,0_2)$ or the second output port, $P(0_1,2_2)$ is independent of $\alpha$.

Fig. \ref{fig:maxp11}(a) depicts the maximal coincidence rate $\textrm{max}_\alpha P(1_1,1_2)$ which occurs at $\alpha = \pi - \frac{\Delta \alpha}{2}, \delta t = 0$ as a function of transmittance $t^2$ and reflectance $r^2$. The cross-sections along the solid lines in Fig. \ref{fig:maxp11}(a) are shown in Fig. \ref{fig:maxp11}(b) in corresponding colors. The cross-sections correspond to different imbalance ratios $t^2/r^2$. A common feature among all the curves is a point of inflexion along the dotted curve and termination on the dashed curve. In the limiting cases of $t^2/r^2 \rightarrow \infty$ or $t^2/r^2 \rightarrow 0$, the two points coincide. The dashed curve corresponds to the coincidence probability in a lossless beam splitter, which varies as $(1-2t^2)^2$. The dotted line corresponds to the coincidence rate at largest value of $t^2$ that allows full programmability, i.e. $\Delta \alpha$ = 2$\pi$. 

\section{Hong-Ou-Mandel like interference}
\begin{figure}[ht]
\centering
\includegraphics[width=8cm]{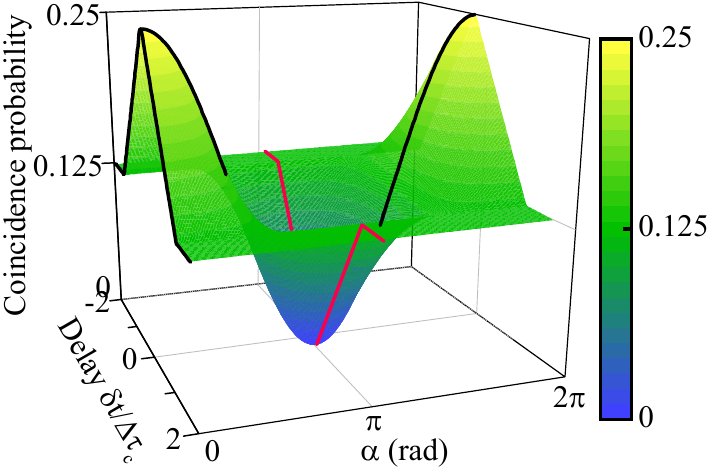}
\caption{The figure depicts the coincidence probability $P(1_1,1_2)$ as a function of delay time ($\delta t$) at various values of $\alpha$ in a lossy symmetric balanced beamspliter with $t = \tau = r = \rho$ = 0.5. The coincidence probability $P(1_1,1_2)$ varies like a cosine with $\alpha$ for perfectly indistinguishable photons $\delta t$ = 0. The conventional Hong-Ou-Mandel dip (red curve) is seen at $\alpha = \pi$ which becomes a peak at $\alpha$ = 0 or 2$\pi$. The triangular shape of the Hong-Ou-Mandel dip or peak is a consequence of the photon pair generation process.}
\label{fig:phasescan}
\end{figure}

In an experiment, the quantum interference can be measured by performing a Hong-Ou-Mandel-like experiment, where the distinguishability of the photons is varied by adding a time delay $\delta t$. Let us suppose that the two photons are generated using collinear type-II spontaneous parametric down conversion in a periodically poled potassium titanyl phosphate (PPKTP) under pulsed pumping (the center frequency and fourier-transformed pulse width of the pump are $\omega_p$ and $\tau_p$ respectively). The resulting bi-photon amplitude of the idler ($\omega_i$) and signal ($\omega_s$) photons is \cite{Grice1997}
\begin{equation}
\psi(\omega_i,\omega_s) = \textrm{sinc}\left( \frac{k_p-k_i-k_s - \frac{2 \pi}{\Lambda}}{\pi}\frac{L}{2} \right) \exp\left\{- \left[(\omega_s+\omega_i-\omega_p)\frac{\tau_p}{2}\right]^2 \right\},
\end{equation}
where $\Lambda$ and $L$ are the poling period and length of crystal, respectively. From the above bi-photon amplitude, the overlap integral $I_\textrm{overlap} (\delta t)$ can be calculated, which gives the coincidence probability $P(1_1,1_2)$. Fig. \ref{fig:phasescan} elucidates the expected Hong-Ou-Mandel-like curve at various values of $\alpha$ for a symmetric balanced beam splitter with $t=r=\rho=\tau=1/2$. The delay time is normalized to the coherence time $\Delta \tau_c$ of the single photons generated by the source. For $\alpha$ = $\pi$, a Hong-Ou-Mandel like dip (red curve) is evident which slowly evolves into a peak as $\alpha$ approaches 0 or 2$\pi$, indicating increased antibunching of photons. The sinusoidal variation of the coincidence probability $P(1_1,1_2)$ for perfectly indistinguishable photons, i.e. $\delta t$ = 0, with the phase $\alpha$ indicates the programmability of quantum interference at these beam splitters. 

\section{Discussion and conclusions}
\begin{figure}[ht]
\centering
\includegraphics[width=8cm]{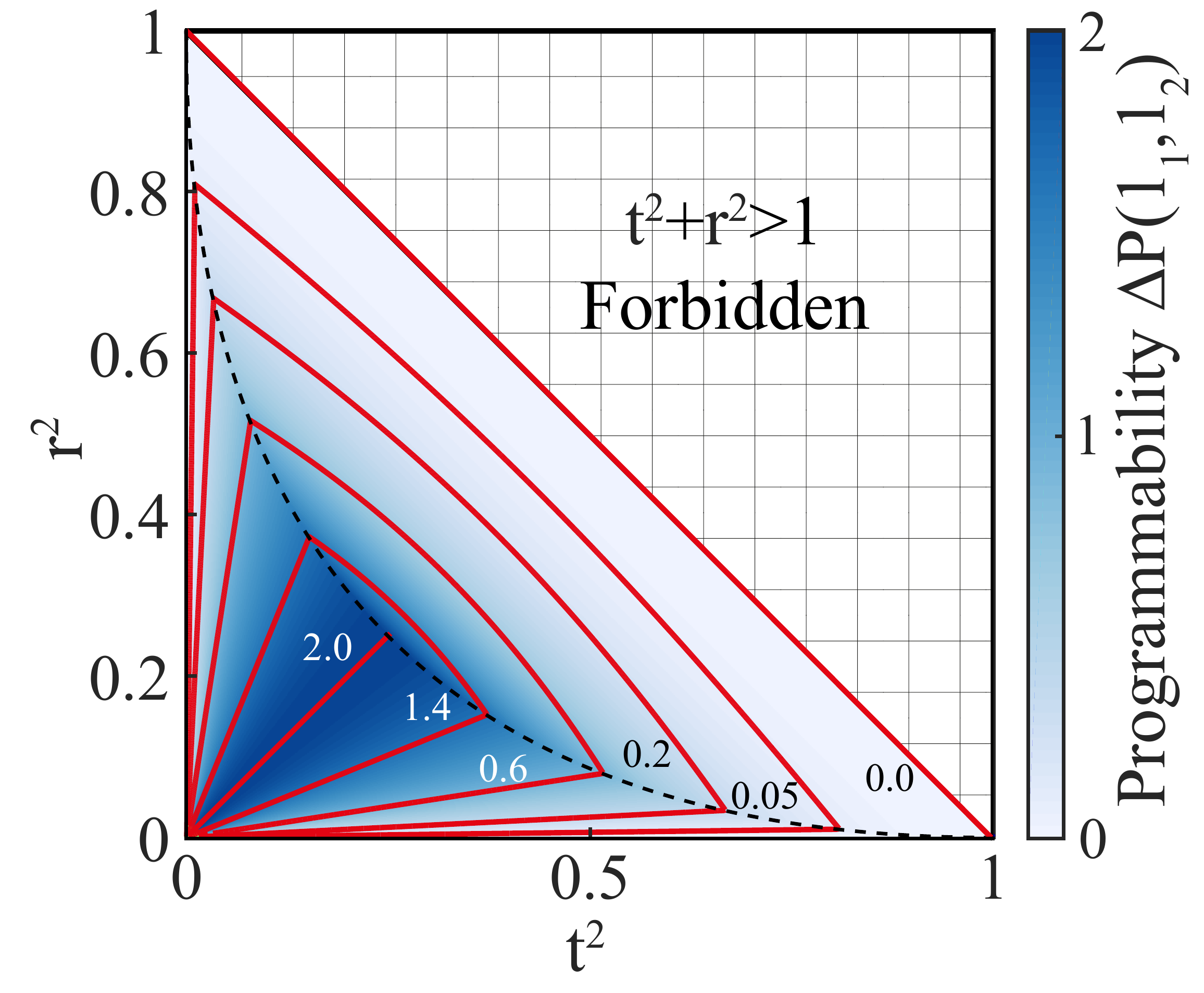}
\caption{Programmability of the coincidence rate $\Delta P(1_1,1_2)$ is depicted here together with few representative contours at values indicated beside them. The black dashed curve represents $t+r$ = 1. The lossless beam splitters have $\Delta P(1_1,1_2)$ = 0, while the balanced lossy beam splitters satisfying $t+r<1$ have maximal programmability with $\Delta P(1_1,1_2)$ = 2.}
\label{fig:tunability}
\end{figure}

Through the above theoretical analysis of a general two-port circuit, we demonstrated that losses introduced in a beam splitter allow the tunability of $\alpha$ and hence of the quantum interference. We can quantify the programmability of quantum interference by defining the parameter $\Delta P(1_1,1_2)$ which is the programmable range of coincidence probability, defined as
\begin{equation}
\Delta P(1_1,1_2) \equiv \frac{\textrm{max}_\alpha P(1_1,1_2) - \textrm{min}_\alpha P(1_1,1_2)}{P(1_1,1_2; \textrm{distinguisable})},
\end{equation}
where, the numerator is the difference between maximum and minimum coincidence probabilities (see Fig. \ref{fig:phasescan}) with indistinguishable photons ($\delta t$ = 0) and the denominator is the coincidence rate with distinguishable photons ($\delta t \rightarrow \pm \infty$). Fig. \ref{fig:tunability} depicts $\Delta P(1_1,1_2)$ as a function of transmittance and reflectance with few representative contours shown in red. The lossless beam splitters, which lie on the diagonal separating the allowed and the forbidden regions, show no programmability. Maximal programmability of $\Delta P(1_1,1_2)$ = 2, is allowed by lossy balanced beam splitters for perfectly indistinguishable photons. The black dashed line in the figure corresponds to $t + r = 1$. While $\Delta \alpha$ = 2$\pi$ in the region $t + r <$ 1, the programmability is not uniform. This arises from the imbalance $t^2/r^2 \neq 1$ in unbalanced beam splitters.

Our theoretical calculations explain the recent experimental demonstrations of programmable quantum interference in opaque scattering media and multimode fibers \cite{Wolterink2016,Defienne2016}. In these experiments, two-port circuits were constructed using wavefront shaping that selects two modes from an underlying large number of modes \cite{Vellekoop2007,Mosk2012}. Light that is not directed into the two selected modes due to imperfect control or noise can be modeled as loss. Typical transmission of $\sim$10\% in opaque scattering media ensures the full programmability when a balanced two-port circuit is constructed \cite{Huisman2014,Huisman2015}.

In summary, we theoretically analyzed the most general passive linear two-port circuit from only energy considerations. We establish the programmability of quantum interference between two single photons in the context of recent experimental demonstrations.

\section*{Acknowledgements}
We would like to thank Klaus Boller, Allard Mosk, and Willem Vos for discussions. The work was financially supported by the Nederlandse Organisatie voor Wetenschappelijk Onderzoek (NWO) and the Stichting Fundamenteel Onderzoek der Materie (FOM).
\end{document}